\documentclass[ reprint, amsmath,amssymb, aps,pra,floatfix,]{revtex4-1}
\pdfoutput=1
\usepackage{graphicx}
\usepackage{float}
\usepackage{wrapfig}
\usepackage{subcaption}
\usepackage{dcolumn}
\usepackage{bm}
\usepackage{hyperref}
\usepackage{natbib}
\begin{document}
\title{Singularity-free quantum tracking control of molecular rotor orientation}
\author{Alicia Magann}
\email{amagann@princeton.edu}
\author{Tak-San Ho}
\email{tsho@princeton.edu}
\author{Herschel Rabitz}
\email{hrabitz@princeton.edu}
\affiliation{%
Department of Chemistry, Princeton University, Princeton, New Jersey 08544, USA\\
}%
\date{\today}
\begin{abstract}
Quantum tracking control aims to identify applied fields to steer the expectation values of particular observables along desired paths in time. The associated temporal fields can be identified by inverting the underlying dynamical equations for the observables. However, fields found in this manner are often plagued by undesirable singularities. In this paper we consider a planar molecular rotor, and derive \emph{singularity-free} tracking expressions for the fields that steer the expectation of the orientation of the rotor along desired trajectories in time. Simulations are presented that utilize two orthogonal control electric fields to drive the orientation of the rotor along a series of designated tracks.
\end{abstract}
\maketitle

\section{Introduction}

The desire to manipulate quantum dynamics has inspired significant research activity for many years \cite{Brif2011,Glaser2015,0953-8984-28-21-213001}. One longstanding goal is to identify control fields capable of driving a quantum system from its initial state to a desired target state at a designated time $t=T$. Such goals have led to the formulation of quantum optimal control theory \cite{Peirce1988}, which has been utilized in many applications including high harmonic generation \cite{RevModPhys.80.117}, quantum information science \cite{Dolde2014,Waldherr2014}, and chemical reactions \cite{Tannor1985,Brumer1986, Abrashkevich1998}. Quantum optimal control seeks fields to steer a system to a target objective using iterative optimization methods \cite{Rothman2005,PhysRevA.84.022326,doi:10.1063/1.1564043,doi:10.1063/1.476575}, which are traditionally carried out with no specific regard for the intervening dynamics linking the initial state to the final target. 

An alternative approach is quantum tracking control, which aims to steer a quantum system from its initial state to a target following a prescribed time-dependent path for the intervening dynamics. Quantum tracking control typically involves two stages: first specifying a trajectory $\langle O(t)\rangle_d$, $\forall t\in[0,T]$, to describe the ``designated'' time-evolution of an operator expectation value $\langle O(t)\rangle\equiv \langle\psi(t)|O|\psi(t)\rangle$ (here, $|\psi(t)\rangle$ is the wave function of the quantum system at time $t$ and $O$ is the observable of interest), and then seeking a control field $\varepsilon(t)$ to track the specified trajectory such that $\langle O(t)\rangle = \langle O(t)\rangle_d$ within the time span $[0,T]$. Numerical studies for quantum tracking control have been carried out in systems ranging from a qubit \cite{Lidar2004} to diatomic and triatomic molecules \cite{Gross1993,Chen1995,Chen1997}, and it has also been accomplished in combination with Lyapunov \cite{Mirrahimi2005, Coron2009, Sahebi2018} and adaptive \cite{Zhu2003} methods. 

The roots of quantum tracking control lie in the engineering literature, beginning with the case of linear control systems \cite{BROCKETT1965548}. Later, many of these key results were generalized for nonlinear control systems \cite{Hirschorn1979} including bilinear quantum control systems \cite{Ong1984}. However, it was found that attempting to \emph{exactly} track arbitrary observable trajectories in nonlinear control systems can lead to \emph{singularities} in the controls, where the control becomes unbounded (i.e., often swinging from a positive to a negative unbounded value when passing through the singularity) and the ability to follow the designated track can break down \cite{Hirschornf1987}. 

In the quantum tracking control formulation, a control field $\varepsilon(t)$ is determined via the solution of an inverse dynamical expression, which is computationally attractive when compared with the arduous iterative optimization methods called for in quantum optimal control. The primary obstacle in the implementation of quantum tracking control is the possible impending field singularities mentioned above \cite{Zhu1905}, which often appear as an artifact of attempting to force a system to evolve along a track inconsistent with its natural dynamics. If singularities can be avoided, however, tracking control can become an efficient means for realizing quantum system objectives. Thus far, there remains no \emph{a priori} approach for prescribing a smooth path $\langle O(t)\rangle_d$ connecting arbitrary initial and final objective values ($\langle O(0)\rangle$ and $\langle O(T)\rangle$, respectively) that assures a well-behaved control field and further investigation of quantum tracking control is needed in order to better assess its general practical utility.

In this paper, we take a step in the latter direction by showing that quantum tracking control of the expectation value of molecular rotor orientation is singularity-free. We illustrate this concept with a series of simulations using two orthogonal, linearly polarized control fields to steer the orientation of a planar rigid rotor along designated trajectories in time. Although the global controllability of rotors subject to two orthogonal control fields has been studied \cite{Turinici2010}, the corresponding tracking control problem has not been considered. Moreover, the control of molecular orientation is important for a number applications including chemical reactions \cite{Stapelfeldt2003} and high harmonic generation \cite{RevModPhys.80.117}. In the laboratory, a planar molecular rotor system could be constructed using laser and evaporative cooling techniques to generate an ultracold molecule, and then adsorbing the molecule onto a surface or trapping it in an ``optical lattice'' created using the interference of optical laser beams \cite{Baranov2012}. Shaped microwave tracking fields can be created experimentally by modulating the field in the time-domain using an arbitrary waveform generator \cite{Yao2011,Lin2005}.

The remainder of the paper is organized as follows. In Section 2 we review the theoretical foundations of quantum tracking control. We then give the model used to describe the planar rotor, followed by the derivation of the tracking control equations to control the orientation of the rotor, highlighting the singularity-free character of these equations. In Section 3 we present a series of numerical examples illustrating the capability of singularity-free tracking of rotor orientation, and in Section 4 we finish with conclusions. 

\section{Theoretical foundations}

\subsection{Quantum tracking control formulation}

Consider a quantum system with a Hamiltonian of the form $H(t) = H_0-\mu\varepsilon(t)$, where $H_0$ is the field-free Hamiltonian, $\varepsilon(t)$ is an applied field, and $\mu$ is the system's dipole operator; in this initial presentation of tracking control principles, $\varepsilon(t)$ is considered to be aligned with $\mu$, while for the planar rotors in Section \ref{section2.3} this restriction is relaxed. The evolution of the expectation value of the observable operator $O$ is governed by the equation
\begin{equation}
\frac{d\langle O(t)\rangle}{dt} = \frac{i}{\hbar}\langle\psi(t)|\big[H_0-\mu\varepsilon(t),O\big]|\psi(t)\rangle
\label{Ehrenfest}
\end{equation}
where $O$ is a time-independent Hermitian operator and $|\psi(t)\rangle$ the state of the quantum system at time $t$. In the quantum tracking control formulation, a designated trajectory $\langle O(t)\rangle_d$, $t\in[0,T]$ is first specified \emph{a priori} for the expectation value $\langle O(t)\rangle$. Then, by assuming $[\mu,O]\neq 0$ and invoking Eq. (\ref{Ehrenfest}), the tracking control field $\varepsilon(t)$, given the trajectory $\langle O(t)\rangle_d$ , can be directly computed as
\begin{equation}
\varepsilon(|\psi(t)\rangle, t) = \frac{i\hbar\frac{d\langle O(t)\rangle_d}{dt}+\langle\psi(t)|[H_0,O]|\psi(t)\rangle}{\langle\psi(t)|[\mu,O]|\psi(t)\rangle}\, .
\end{equation}
In the situation where $[\mu,O]\equiv 0$, additional time derivatives of Eq. (\ref{Ehrenfest}) need to be taken until $\varepsilon(t)$ appears explicitly. In general, for $k$ additional time derivatives, with the simplified notation $O_{k}\equiv \frac{i}{\hbar}[H_0,O_{k-1}]-\frac{i}{\hbar}[\mu,O_{k-1}]\varepsilon(t)$, where $O_0=O$ and $k=1,\cdots$, and assuming $[\mu,O_k]\not\equiv 0$ we then obtain a working expression of the following form \cite{Gross1993},
\begin{equation}
\varepsilon(|\psi(t)\rangle,t) = \frac{i\hbar\frac{d^{k+1}\langle O(t)\rangle_d}{dt^{k+1}}+\langle\psi(t)| [H_0,O_k]|\psi(t)\rangle }{\langle \psi(t)|[\mu,O_k]|\psi(t)\rangle}\, ,
\label{epsilon}
\end{equation}
in which the denominator $\langle\psi(t)|[\mu,O_k]|\psi(t)\rangle$ is generally nonzero, but may still pass through some isolated zeros and change sign, thus causing $\varepsilon(t)$ to possess singularities. To use Eq. (\ref{epsilon}) to compute the tracking control field, $k$ initial conditions $\langle O(t=0)\rangle$, $\frac{d\langle O(t=0)\rangle}{dt}$, $\cdots$, $\frac{d^{k}\langle O(t=0)\rangle}{dt^k}$ are needed to ensure consistency with the designated track $\langle O(t)\rangle_d$.

The underlying time-dependent Schr\"odinger equation  
\begin{equation}
i\hbar\frac{\partial}{\partial t}|\psi(t)\rangle = \big(H_0-\mu\varepsilon(|\psi(t)\rangle,t)\big)|\psi(t)\rangle
\label{nonlin}
\end{equation}
is highly nonlinear due to the functional dependence of the field $\varepsilon(|\psi(t)\rangle,t)\big)$ on $|\psi(t)\rangle$. In practice, the coupled Eqs. (\ref{epsilon}) and (\ref{nonlin}) may be solved in the following fashion. We start with the initial field value $\varepsilon(t=0)$, which can be obtained by evaluating Eq. (\ref{epsilon}) using the initial condition for $|\psi(0)\rangle$ and any necessary derivatives at $t=0$. We then propagate the system forward by integrating Eq. (\ref{nonlin}) over a small time step $|\psi(t=0)\rangle\rightarrow|\psi(t=\Delta t)\rangle$. The updated system state $|\psi(\Delta t)\rangle$ is then substituted back into Eq. (\ref{epsilon}), which is followed by another propagation, i.e., $|\psi(\Delta t)\rangle\rightarrow|\psi(2\Delta t)\rangle$, and the same process is repeated until the target time is reached or a singularity is encountered. If the latter circumstance arises, various methods have been suggested to deal with the situation \cite{Zhu2003,Mirrahimi2005, Coron2009, Sahebi2018}, but a fully satisfactory general procedure is yet to be found.

\subsection{Planar rigid rotor model}

We consider a linear rigid rotor in a plane with dipole moment vector $\vec{\mu}(\varphi)$, where $\varphi$ denotes the rotational angle of the dipole moment with respect to the $\hat{x}$-axis. The $\hat{x}$ and $\hat{y}$ projections of the rotor's dipole vector are given by $\mu_x(\varphi) = \mu\cos\varphi$ and $\mu_y(\varphi) = \mu\sin\varphi$ respectively. The rotor is coupled through the dipole moment to two orthogonal control fields $\varepsilon_x(t)$ and $\varepsilon_y(t)$, linearly polarized along the $\hat{x}$ and $\hat{y}$ axes, respectively. The rotor's Hamiltonian is given by
\begin{equation}
\begin{aligned}
H(\varphi,t)=-B\frac{\partial^2}{\partial\varphi^2}-\mu\varepsilon_x(t)\cos\varphi-\mu\varepsilon_y(t)\sin\varphi \, ,
\label{eq:Hamiltonian}
\end{aligned}
\end{equation}
where $B=\frac{\hbar^2}{2I}$ is the rotational constant, $\hbar$ is the reduced Planck's constant, and $I$ is the rotor's moment of inertia. 

The rotor is studied in the basis of the eigenstates $|m\rangle$ of the angular momentum operator, $L^2=-\hbar^2\frac{\partial^2}{\partial\varphi^2}$, satisfying the eigenvalue equation,
\begin{equation}
\begin{aligned}
L^2|m\rangle=m^2\hbar^2|m\rangle \, ,
\end{aligned}
\end{equation}
where $m=-M,-M+1,...,-1,0,1,...M-1,M$. The eigenstates $|m\rangle$ can be expanded as:
\begin{equation}
|m\rangle=\int_0^{2\pi}|\varphi\rangle\langle\varphi|m\rangle d\varphi
\end{equation}
in terms of the angle $\varphi\in[0,2\pi]$ where 
\begin{equation}
\begin{aligned}
\langle\varphi|m\rangle=\sqrt{\frac{1}{2\pi}}e^{im\varphi}\, ,
\end{aligned}
\end{equation}
noting that $\int_0^{2\pi}|\varphi\rangle\langle\varphi|d\varphi=1$. In this basis, the angular terms in Eq. (\ref{eq:Hamiltonian}) can be calculated using the matrix element relations,
\begin{equation}
\begin{aligned}
\langle m|\cos\varphi|m'\rangle&=\frac{1}{2}\big\{\delta_{m,m'+1}+\delta_{m,m'-1}\big\}\\
\langle m|\sin\varphi|m'\rangle&=\frac{-i}{2}\big\{\delta_{m,m'+1}-\delta_{m,m'-1}\big\}\, .
\end{aligned}
\end{equation}
The dynamics of the rotor are governed by the time-dependent Schr\"odinger equation,
\begin{equation}
i\hbar\frac{\partial}{\partial t}|\psi(t)\rangle=H(\varphi,t)|\psi(t)\rangle,\quad|\psi(0)\rangle=|\psi_0\rangle\, ,
\label{SchrodingerEq}
\end{equation}
where $H(\varphi,t)$ is given in Eq. (\ref{eq:Hamiltonian}).

\subsection{Simultaneous tracking of orientation observables}
\label{section2.3}

\begin{figure}[t]
\centering
\includegraphics[scale=0.2]{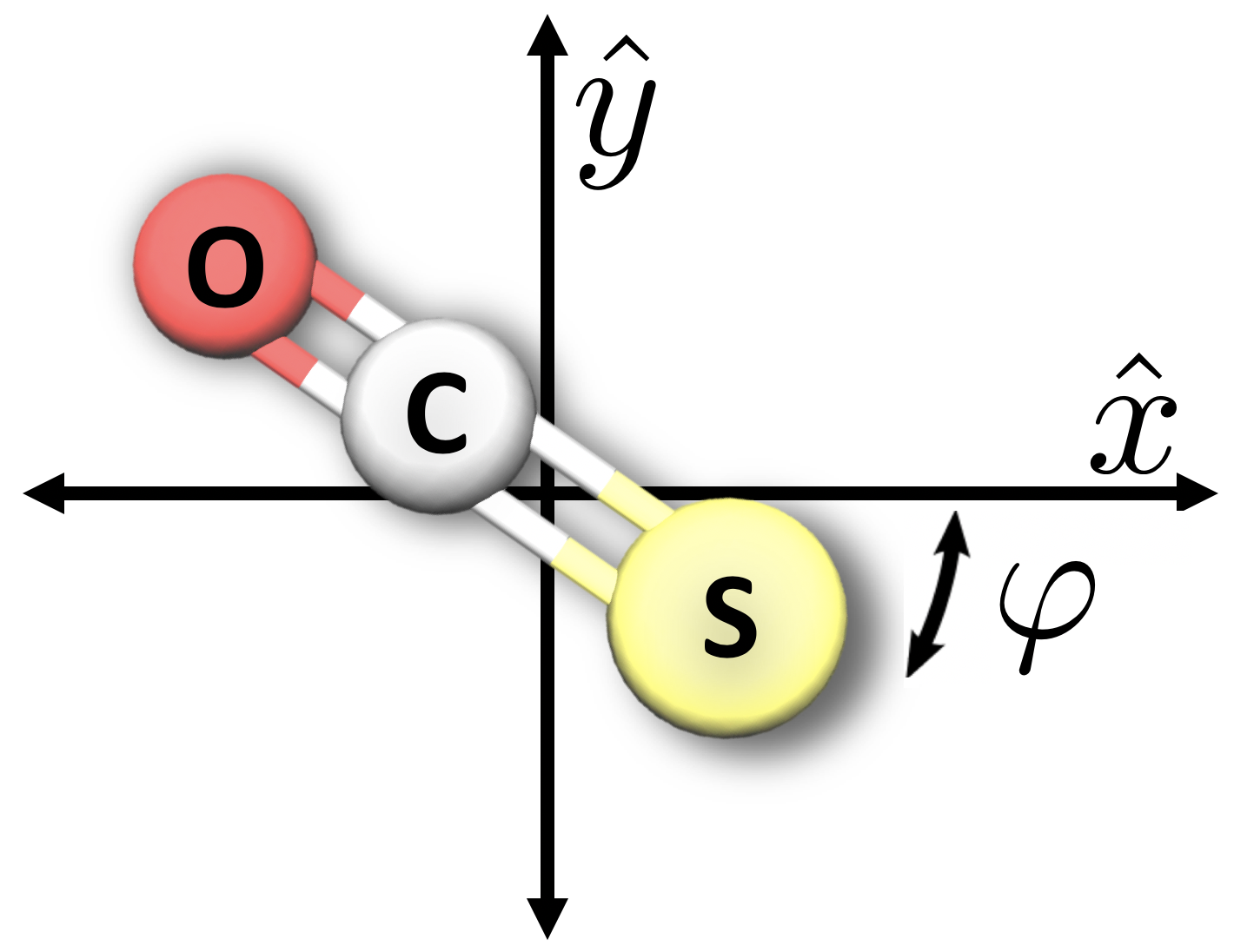}
\caption{Diagram of planar molecular rotor considered in this work, where the chemical symbol labels O, C, and S, respectively, denote oxygen, carbon, and sulfur. The rotor's center of mass is shifted towards the sulfur bond.}
\label{rotorfigure}
\end{figure}

Here we derive equations for computing the orthogonal fields $\varepsilon_x(t)$ and $\varepsilon_y(t)$ that simultaneously track the expectation values of the rotor orientation observables. These observables are defined as the $\hat{x}$ and $\hat{y}$ projections of the rotor's dipole vector (normalized with respect to $\mu$), $O_x \equiv\cos\varphi$ and $ O_y\equiv \sin\varphi$ respectively. We remark that the singularity-free nature of tracking atomic and molecular dipoles outside of an orientation context has been studied in \cite{Campos2017}. 

The tracking equation requires going to second order to determine the fields, and the resultant coupled differential equations governing $\langle O_x (t) \rangle$ and $\langle O_y(t)\rangle$ are given by
\begin{equation}
\begin{aligned}
\frac{d^2\langle O_x(t)\rangle}{dt^2}& =\frac{-1}{\hbar^2}\langle\psi(t)|\Big[H(\varphi,t),\big[H(\varphi,t),\cos\varphi\big]\Big]|\psi(t)\rangle\, ,\\
\frac{d^2\langle O_y(t)\rangle}{dt^2}& =\frac{-1}{\hbar^2}\langle\psi(t)|\Big[H(\varphi,t),\big[H(\varphi,t),\sin\varphi\big]\Big]|\psi(t)\rangle\, .
\label{Eq:dynamical}
\end{aligned}
\end{equation}
where $H(\varphi,t)$ is given in Eq. (\ref{eq:Hamiltonian}). Eqs. (\ref{Eq:dynamical}) are two coupled \emph{algebraic} equations in the fields $\varepsilon_x(t)$ and $\varepsilon_y(t)$. Thus, after rearranging them and substituting in the designated tracks $\langle O_x(t)\rangle = \langle O_x(t)\rangle_d$ and $\langle O_y(t)\rangle = \langle O_y(t)\rangle_d$, the equations can simultaneously be solved for $\varepsilon_x(t)$ and $\varepsilon_y(t)$ to give

\begin{widetext}
\begin{equation}
\begin{aligned}
\left( \begin{array}{c} \varepsilon_x(t) \\ \varepsilon_y(t) \end{array} \right) = \frac{1}{D(\varphi,t)}\frac{2\mu B}{\hbar^2}& \begin{pmatrix} \langle\psi(t)|\cos^2\varphi|\psi(t)\rangle & \langle\psi(t)|\cos\varphi\sin\varphi|\psi(t)\rangle \\ \langle\psi(t)|\sin\varphi\cos\varphi|\psi(t)\rangle &\langle\psi(t)|\sin^2\varphi|\psi(t)\rangle \end{pmatrix} \\
&\text{   }\times \left( \begin{array}{c} \frac{d^2\langle O_x(t)\rangle_d}{dt^2}+\frac{B^2}{\hbar^2}\langle\psi(t)|\cos\varphi+4\sin\varphi\frac{\partial}{\partial\varphi}-4\cos\varphi\frac{\partial^2}{\partial\varphi^2}|\psi(t)\rangle \\ \frac{d^2\langle O_y(t)\rangle_d}{dt^2}+\frac{B^2}{\hbar^2}\langle\psi(t)|\sin\varphi-4\cos\varphi\frac{\partial}{\partial\varphi}-4\sin\varphi\frac{\partial^2}{\partial\varphi^2}|\psi(t)\rangle \end{array} \right)\, ,
\label{fields}
\end{aligned}
\end{equation}
\end{widetext}
where the determinant 
\begin{equation}
\begin{aligned}
D(\varphi,t)=\Big(\langle\psi(t)|&\sin^2\varphi|\psi(t)\rangle\Big)\Big(\langle\psi(t)|\cos^2\varphi|\psi(t)\rangle\Big)\\
&-\Big(\langle\psi(t)|\sin\varphi\cos\varphi|\psi(t)\rangle\Big)^2\geq 0
\label{DeterminantEq}
\end{aligned}
\end{equation}
is positive semidefinite due to the Cauchy-Schwarz inequality between the two vectors $\cos\varphi|\psi(t)\rangle$ and $\sin\varphi|\psi(t)\rangle$ (i.e., $\langle\phi_1|\phi_1\rangle\langle\phi_2|\phi_2\rangle\geq|\langle\phi_1|\phi_2\rangle|^2$ between two arbitrary vectors $|\phi_1\rangle$, $|\phi_2\rangle$). Although this circumstance eliminates the chance that $D(\varphi,t)$ can change sign, singularities could still appear in the rare event of the equality $D(\varphi,t)=0$. To eliminate this possibility as well, the condition
\begin{equation}
\langle \psi(t)|\cos\varphi|\psi(t)\rangle^2 +  \langle\psi(t)|\sin\varphi|\psi(t)\rangle^2 \in (0,1)
\label{condition}
\end{equation} 
must be met, as this condition renders $D(\varphi,t)$ to be strictly positive. As a practical matter, the tracks $\langle O_x(t)\rangle_d$ and $\langle O_y(t)\rangle_d$ must remain within (but not touching the boundaries of) the unit circle; this restriction assures the strictly positive character of $D(\varphi,t)$, and as a result the fields found from Eq. (\ref{fields}) will be smooth and free of singularities.

\section{Numerical illustrations}

In this section we provide three simulation examples that highlight the capability to simultaneously track the $\hat{x}$ and $\hat{y}$ orientations of a linear molecular rotor in a singularity-free manner. In particular, we consider a planar OCS rigid rotor, see Fig. (\ref{rotorfigure}). The magnitude of the dipole moment of OCS is $\mu=0.709$ Debye \cite{PhysRev.77.500} and its rotational constant is $B=0.203 \text{ cm}^{-1}$ \cite{doi:10.1063/1.3253139}. For all simulations, the rotor is initialized in its ground rotational state $|m\rangle=|0\rangle$.

\subsection{Gaussian Tracks}
We begin by showing the utility of the quantum tracking control based on Eq. (\ref{fields}) to track two perpendicular Gaussian trajectories defined as,
\begin{equation}
\begin{aligned}
\langle O_x(t)\rangle_d &= \alpha e^{-\big(\frac{t-0.4T}{T/15}\big)^2}\, ,\\
\langle O_y(t)\rangle_d &=\alpha e^{-\big(\frac{t-0.8T}{T/15}\big)^2}\, ,
\end{aligned}
\label{gaussians}
\end{equation}
along the $\hat{x}$- and $\hat{y}$-axes, respectively, where the pulse length $T = 50\hbar/B$ and $\alpha = 0.9$ (i.e., note that $\alpha$ must satisfy $\alpha < 1$ to ensure consistency with Eq. (\ref{condition})). 

\begin{figure}[t]
\centering
\includegraphics[scale=0.7]{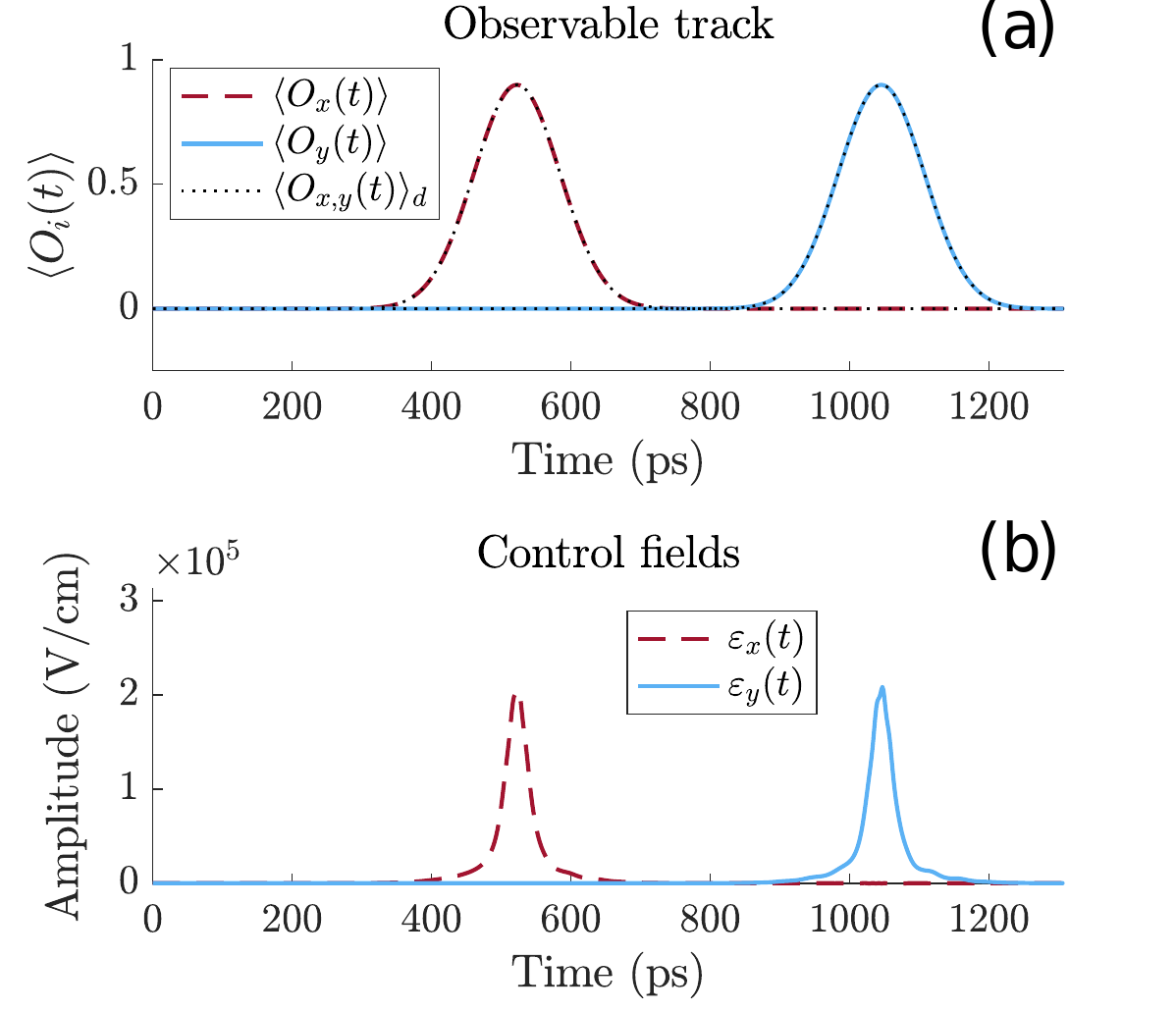}
\caption{(a) The time-evolution of $\langle O_x(t) \rangle$ and $\langle O_y(t) \rangle$ is shown when the $\varepsilon_x(t)$ and $\varepsilon_y(t)$ fields given in (b) are applied to the rotor. The designated tracks $\langle O_x(t) \rangle_d$ and $\langle O_y(t) \rangle_d$ are also shown (black dotted curve), which exactly superimpose with the actual time evolution. }
\label{gauss}
\end{figure}

Fig. \ref{gauss}a shows the two Gaussian profiles given in Eq. (\ref{gaussians}),  well separated by  a time interval $0.4T=522$ ps, which is more than twice their individual full width at half maximum, $\sigma=241$ ps. Fig. \ref{gauss}b shows the two compact tracking control fields that appear in succession to alternately steer $\langle O_x(t) \rangle$ and $\langle O_y(t) \rangle$ along these two Gaussian tracks. The designated tracks are followed, and the peak positions of the tracking fields coincide with those of the respective Gaussian trajectories. Importantly, the control fields are smooth and free of singularities as expected.

\subsection{Spiral Track} \label{SpiralSection}

\begin{figure}[t]
\centering
\includegraphics[scale=0.7]{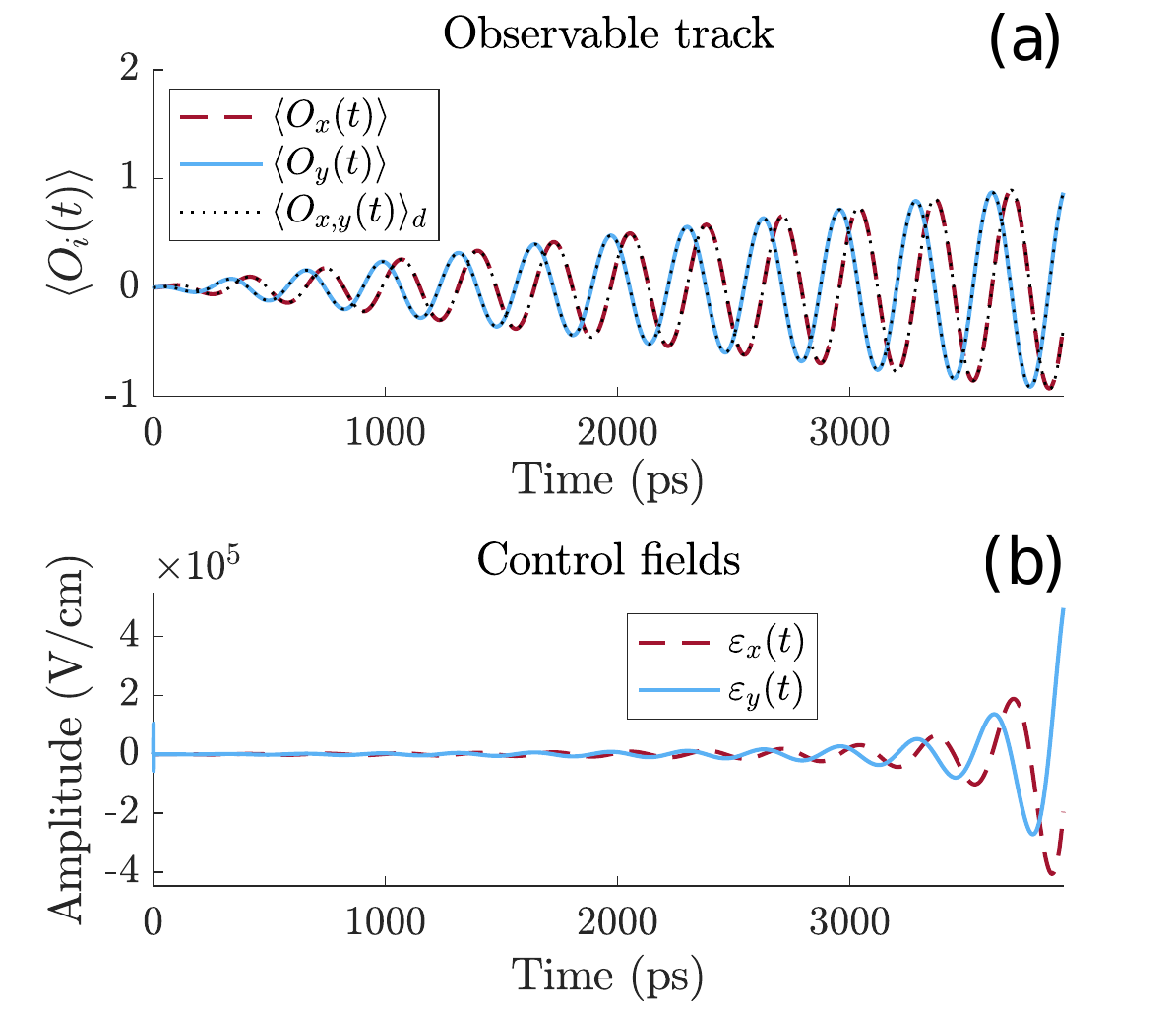}
\caption{(a) The time-evolution of $\langle O_x(t) \rangle$ and $\langle O_y(t) \rangle$ is shown when the $\varepsilon_x(t)$ and $\varepsilon_y(t)$ fields given in (b) are applied to the rotor. The designated tracks $\langle O_x(t) \rangle_d$ and $\langle O_y(t) \rangle_d$ are also shown (black dotted curve), which exactly superimpose with the actual time evolution. }
\label{spiral}
\end{figure}

Here we show that the orientation tracking protocol also enables following complicated curves in the $\hat{x}-\hat{y}$ plane based on following coordinated paths for $\langle O_x(t)\rangle$ and $\langle O_y(t)\rangle$. Below, we consider a spiral function generated by the tracks (see Fig. \ref{spiral} (a)), given by,
\begin{equation}
\begin{aligned}
\langle O_x(t)\rangle_d &= \beta t \sin(\omega t) f(t)\, ,\\
\langle O_y(t)\rangle_d &= \beta t \cos(\omega t) f(t)\, ,
\end{aligned}
\label{spiraleq}
\end{equation}
where $\beta = 0.95/T$ and $\omega = B/2\hbar$ and
\begin{equation}
f(t)=\frac{1}{\big(1+c_1e^{-c_2t}\big)^{1/c_3}}
\end{equation}
with $T = 150\hbar/B$, $c_1 = T/4$, $c_2 = 0.0002$, and $c_3 = 0.2$. In Eq. (\ref{spiraleq}), the sigmoid function $f(t)$ is included to ensure a smooth start for the spiral. The tracks in Eq. (\ref{spiraleq}) terminate at $t=T$ which prevents violation of the condition in Eq. (\ref{condition}).

\begin{widetext}

\begin{figure}
\centering
\begin{subfigure}{.5\textwidth}
  \centering
  \includegraphics[width=1.0\linewidth]{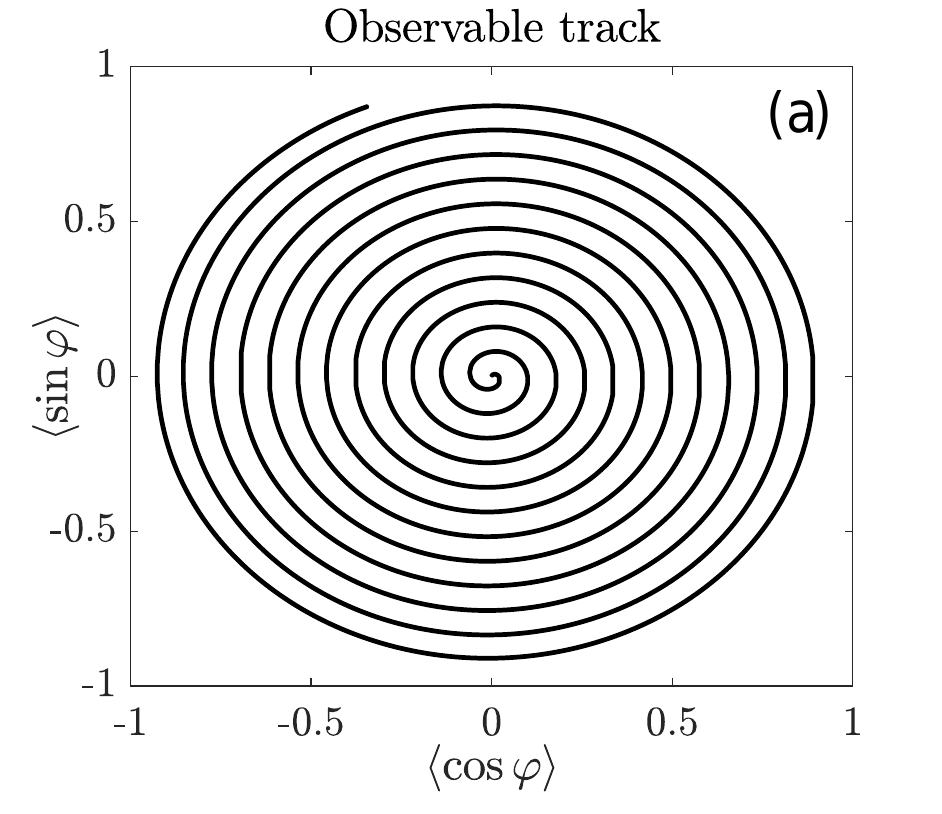}
  \label{fig:sub1}
\end{subfigure}%
\begin{subfigure}{.5\textwidth}
  \centering
  \includegraphics[width=1.0\linewidth]{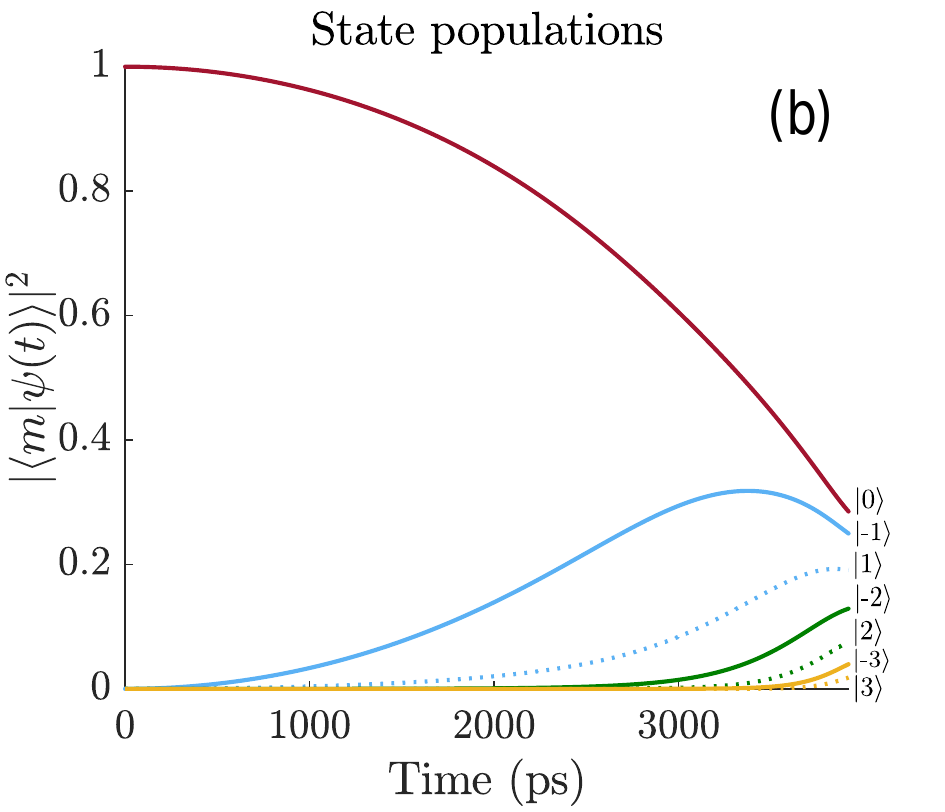} 
  \label{fig:sub2}
\end{subfigure}
\caption{(a) The track followed by the rotor orientation when the fields in Fig. \ref{spiral} (b) are applied, plotted as $\langle O_y(t)\rangle$ versus $\langle O_x(t)\rangle$ in the 2D plane, while (b) shows the time-evolution of the rotational state populations as the track is followed.}
\label{spiral2}
\end{figure}

\end{widetext}

Fig. \ref{spiral} (b) shows that $\varepsilon_x(t)$ and $\varepsilon_y(t)$ oscillate in ever-growing amplitude with increasing time such that the expectation values (red dashed and blue solid curves) track the functions given in Eq. (\ref{spiraleq}) (black dotted curve), indicated by the trajectories given in Fig. \ref{spiral} (a). These two individual tracks are then plotted as $\langle O_y(t)\rangle$ versus $\langle O_x(t)\rangle$ in the 2D plane in Fig. \ref{spiral2} (a), which shows that they trace out a spiral, as designed. Furthermore, we note that as the track spirals outwards, more rotational states become involved in the dynamics, which is illustrated in Fig. \ref{spiral2} (b). The behavior likely arises as $\langle\varphi|\psi(t)\rangle$ needs to become an ever narrower wave packet in $\varphi$ when the spiral approaches its boundary limits of the unit circle. This increasing involvement of higher rotational states also occurs (not shown here) when approaching the apex of the Gaussian tracks in Section 3.1; when each track later slopes downwards, the number of states involved decreases.

\subsection{Tracking Cursive Script}

\begin{figure}
\centering
\includegraphics[scale=0.7]{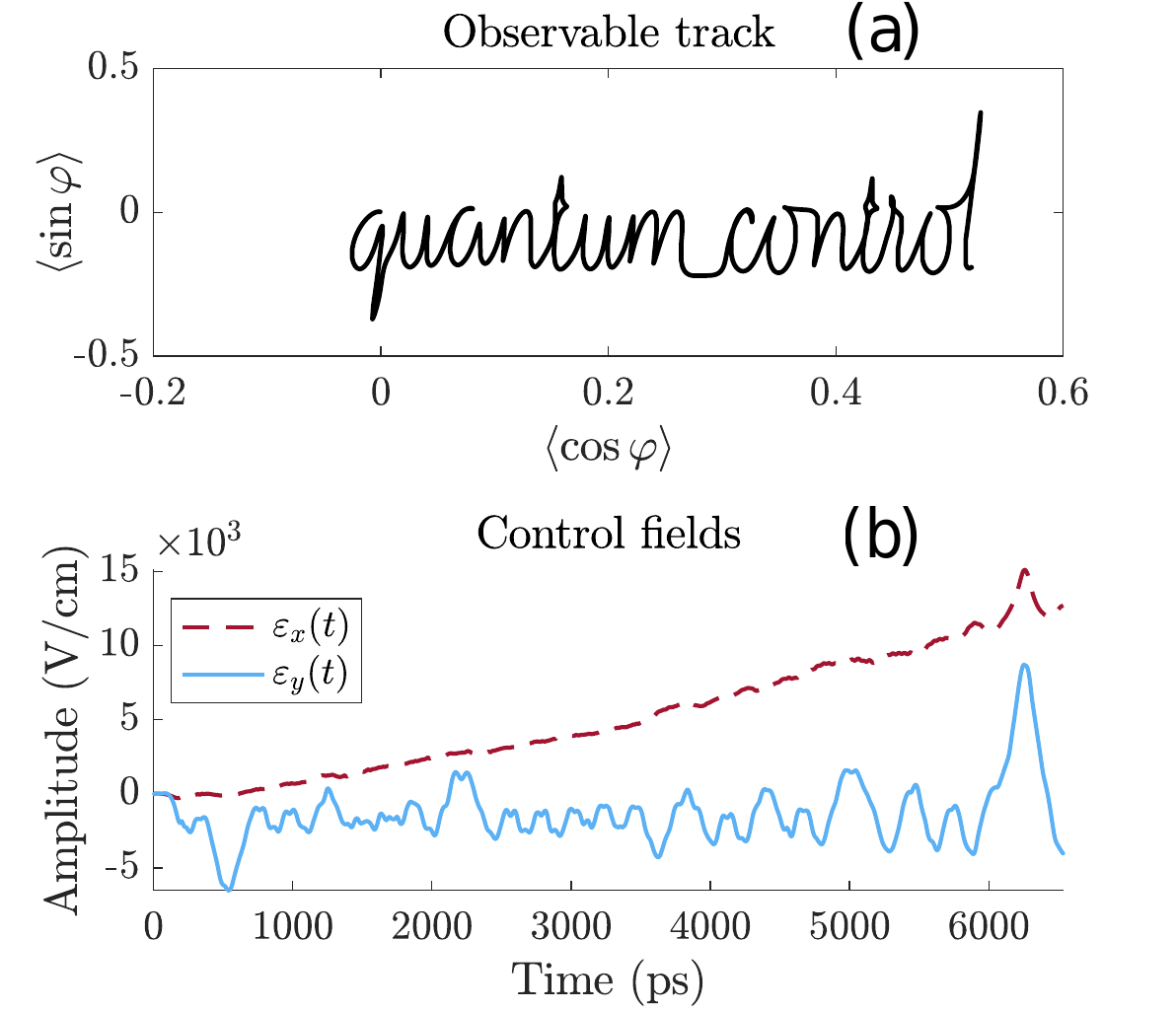}
\caption{(a) shows the track plotted as $\langle O_y(t) \rangle$ versus $\langle O_x(t)\rangle$ in the 2D plane, while (b) shows the corresponding $\hat{x}$ and $\hat{y}$ tracking control fields.}
\label{shield2}
\end{figure}

The spiral example given in Section \ref{SpiralSection} illustrates the capability to trace out curves in the $\langle O_y(t)\rangle$ versus $\langle O_x(t)\rangle$ 2D plane (naturally bounded to remain within the unit circle). To further explore the flexibility of orientation tracking, we conclude by considering a particularly complicated track created from the words ``quantum control'' written in cursive script, see Fig. \ref{shield2} (a), rather than an analytical function. To form the tracks, the outline of the scripted words was first hand-digitized, generating a data set of 812 $(\hat{x},\hat{y})$ coordinates. Additional coordinates were subsequently added by interpolating between these points. The $\hat{x}$ and $\hat{y}$ coordinates were then time-ordered and utilized as forming the $\langle O_x(t)\rangle_d$ and $\langle O_y(t)\rangle_d$ tracks, respectively, such that following the track corresponds to tracing out the words. When these tracks are substituted into Eq. (\ref{fields}), the resultant tracking fields are capable of precisely tracing out the cursive words successfully, see Fig. \ref{shield2} (a) and (b). For a movie of this tracking in time, we refer the interested reader to the online supplementary material \cite{SuppMat}.

In the current example the data tracking scheme is successful, although we remark that it can still lead to very noisy control fields, depending on the nature of the data set. As such, rather than applying the raw fields generated from solving Eq. (\ref{fields}), all of the results shown in Fig. \ref{shield2} have been obtained after filtering the high frequency components out of the raw fields apparently arising from the digitization of the scripted words. However, it was found that the filtering had a very minimal effect on the field's ability to steer the rotor along the desired track (i.e., the deviations are not visibly evident in Fig. \ref{shield2}a between the original track of the words and that achieved numerically). 

\section{Conclusion}

We have demonstrated that quantum tracking control of the orientation of a single planar molecular rotor is singularity-free. The coupled tracking control equations in Eq. (\ref{fields}) that can be solved to produce orthogonal fields $\varepsilon_x(t)$ and $\varepsilon_y(t)$ capable of steering $\langle\psi(t)|\cos\varphi|\psi(t)\rangle$ and $\langle\psi(t)|\sin\varphi|\psi(t)\rangle$ along pre-specified, otherwise arbitrary trajectories. These coupled equations were shown to yield singularity-free fields due to the positive definite character of the determinant (see Eqs. (\ref{DeterminantEq}) and (\ref{condition})). Although we have only considered a single planar rotor, the formulation presented here can be extended to simultaneously track the orientations of multiple dipole-dipole coupled rotors as well, and it can be shown that this will likewise yield fields that are singularity-free.

For illustrations in this work, we have successfully studied various tracking control scenarios for the orientation of a rotor, including tracking smooth Gaussian functions, a parameterized spiral function, as well as rather arbitrary scripted curves, among other cases (not shown here). Finally, it is important to point out that such tracks can drive the rotor to a maximally oriented state, for example, the Gaussian track provides a monotonic rise to the maximal value, whereas the spiral function rotates towards that limit. 

We remark that although the tracking control methodology presented in this paper leads to singularity-free tracking, it does not eliminate the need to define physically acceptable tracks. For example, consideration needs to be given to the rotor's natural time scales (e.g. moving the track too fast near some time $\tau$ could lead to a loss of local controllability in the neighborhood of $\tau$ and resultant track deviations), as well as inherent constraints in the system (e.g., remaining within the unit circle in $\langle\psi(t)|\sin\varphi|\psi(t)\rangle$ versus $\langle\psi(t)|\cos\varphi|\psi(t)\rangle$ space). Another numerical issue is the sensitivity of Eq. (\ref{fields}) with respect to the size of the steps taken in time. In our work, we found that in many cases very small time steps were required in order to achieve the desired tracking; similar demands can arise in many optimal control simulations.

The work presented in this paper is only one step towards the larger goal of developing a fully general tracking control procedure yielding smooth, singularity-free fields for arbitrary observables. To the best of our knowledge, although various tracking control schemes have been suggested for overcoming or bypassing the singularities, so far no fully attractive general procedure has been found. Hopefully further research will lead to singularity-free tracking control, applicable to arbitrary quantum systems. 

{\em Acknowledgements.} -- We thank Linhan Chen for many useful discussions. A.M. acknowledges support from the DOE CSGF. T.S.H. acknowledges support from the DOE-DE-FG02-02ER15344 and H.R. from the ARO W911NF-16-1-0014. 

\bibliography{bib}





   


\end{document}